\documentclass{PoS}
\usepackage{epsfig}

\title{Theoretical progress on cusp effect and Kl4 decays}

\ShortTitle{Theoretical progress on cusp effect and Kl4 decays}

\author{\speaker{J\"urg Gasser}\,\footnote{
Work supported by the Swiss National Science Foundation, and by
the EU contract N0. MRTN--CT--2006--035482,''FLAVIAnet''.}\\
        Inst. Theor. Physik\\ Universit\"at Bern\\ Sidlerstrasse 5\\ Ch--3012 Bern\\
        E--mail: \email{gasser@itp.unibe.ch}}

\abstract{The cusp effect in $K\to 3\pi$ and data on  $K_{e4}$ decays allow one to extract 
 experimental information on the elastic  $\pi\pi$ scattering amplitude near threshold,
 and to confront the outcome of the analyses  with  predictions
  made in the framework of QCD. In my talk I pointed out that these predictions concern an isospin
 symmetric   world,  while experiments are carried out in the real world, where isospin
 breaking effects -- generated by electromagnetic 
interactions and by   the mass difference of the up and down 
quarks --  are always present. 
 I discussed the corrections required to account for these, so that a meaningful comparison 
with the predictions becomes possible. In particular, I pointed out that there is a 
 spectacular isospin breaking effect in $K_{e4}$ decays, overlooked 
 so far. Once it is taken into account,
 the previous
discrepancy between NA48/2 data on $K_{e4}$ decays and the  prediction  of $\pi\pi$ scattering lengths 
disappears.}

\FullConference{Kaon International Conference\\
		 May 21--25, 2007\\
		 Laboratori Nazionali di Frascati dell'INFN}

\begin{document}
\newcommand{\bea}{\begin{eqnarray*}}
\newcommand{\eea}{\end{eqnarray*}}
\newcommand{\bean}{\begin{eqnarray}}
\newcommand{\eean}{\end{eqnarray}}
\newcommand{\nnnl}{\nonumber\\}
\newcommand{\ed}{\end{document}}
\newcommand{\scs}{\,,\,}
\newcommand{\fs}{\, .\,}

\section{Introduction}
Chiral perturbation theory (ChPT) \cite{weinberg79}, combined with Roy equations,  allows one 
to make very precise predictions
for the values of the threshold parameters in elastic $\pi\pi$ 
scattering \cite{scattnpb} -- see Colangelo's 
contribution to these proceedings for a status report \cite{colangelokaon07}.
 Several experiments allow one
 to confront these predictions with the data: i) $K^+\to \pi^+\pi^- e^+\nu_e$ decays
\cite{ke4,blochkaon07,dilella}, 
 ii) the pionium lifetime, measured by the DIRAC collaboration  \cite{tauscher}, and iii) the 
cusp effect in $K\to 3\pi$ decays, investigated by the NA48/2 collaboration  \cite{dilella}.

The experiments performed by the NA48/2 collaboration have generated an impressive data basis, 
as a result of which the  matrix elements of $K_{e4}$ and $K\to 3\pi$ decays 
can be determined with an unprecedented accuracy. The interpretation of these measurements 
was the topic of my talk. In particular, I pointed out that the theoretical 
predictions and the measurements are performed 
in two different settings: the predictions concern pure QCD, in the isospin 
symmetry limit $m_u=m_d$, with photons absent.
 To be more precise, the convention is to choose the quark masses and the 
renormalization group invariant scale of QCD such that
the pion and the kaon masses coincide with the values of the charged ones, and the pion 
decay constant is $F_\pi=92.4$ MeV. 
[I do not specify the masses of the heavy quarks, because in the present context,
 their precise values do not matter.] I refer to this framework as a {\it paradise world}.

On the other hand, experiments are all carried out in the presence of isospin breaking effects, 
generated by real and virtual photons, and by the mass difference of the up and down quarks: this is the 
{\it real world}, described by the Standard Model.
 We are thus faced with the problem to find the relation between 
 quantities measured in the real world, where isospin breaking effects are always present,
and the predictions made  in the paradise world. 

{\underline{Remarks:} }The situation is quite involved: 
1. The above mentioned three processes do not  take place in the 
paradise world: the kaon is stable, whereas  pionium cannot form, because
 electromagnetic interactions are absent.
2. In the real world, the processes i) and iii) do  not occur either, 
 because decays that involve external charged particles 
 have zero probability to happen without the emission of photons. 3.
Lattice calculations
 are not confronted with this problem: the evaluation of e.g. $\pi\pi$ or $\pi K$ scattering lengths 
\cite{scatt_lattice}
 can be performed directly in the framework of 
pure QCD, at $m_u=m_d$: isospin breaking effects can  be excluded  by fiat. 
I refer the interested reader to Colangelo's contribution  \cite{colangelokaon07} , and to Ref.~\cite{leutlatt}
 for a short review of lattice results concerning $\pi\pi$ scattering lengths.

\section{The cusp in $K\to 3\pi$}
Cabibbo and Isidori pointed out \cite{cusp1} that cusp effects in the decay $K^+\to \pi^+\pi^0\pi^0$ may allow 
one to measure  $S$-wave $\pi\pi$ scattering 
lengths,  and thus to check the precise predictions made in the framework of chiral 
perturbation theory, see also \cite{cusp2}. 
A first data analysis based on this observation has appeared \cite{expcusp}
 already some time ago, and the latest developments were discussed  at this conference
 \cite{dilella}.

We have started to investigate this process in the 
framework of non--relativistic quantum field theory in Ref.~\cite{cuspwe}. 
The  status of our calculation is as follows.
\begin{itemize}
\item
Our amplitude  agrees with the one of \cite{cusp1,cusp2} 
  near the cusp, whereas it differs away from it. 
The NA48/2 collaboration at CERN has analyzed data with our amplitude as well. 
The outcome for the scattering length combination  $a_0-a_2$ is in agreement \cite{dilellaprivate}
with the use of the amplitude from \cite{cusp1}.
\item
We have  evaluated  the decay amplitude for the neutral 
kaon, $K_L^0\to \pi^+\pi^-\pi^0, \pi^0\pi^0\pi^0$, 
have written a pertinent FORTRAN code  for the decay probability
and made it available to the NA48/2 collaboration \cite{cusprad}.
In the channel with 3 neutral pions, a cusp occurs as well. In this case, its strength 
is suppressed  \cite{dilella}. Nevertheless, the huge amount of 
data available at NA48/2 should make it possible to extract the same combination of 
scattering lengths as well, although with less precision \cite{dilella}.
\item
It remains to calculate electromagnetic corrections to these decays. 
Because we make use of a quantum field theory framework, our method allows 
us to perform these calculations in a straightforward, yet tedious manner. 
We have investigated the role of the formation of pionium near the cusp and have 
identified an interval around the cusp, where radiative corrections can be calculated 
reliably, in a systematic manner \cite{cusprad}. 
 [In Ref.~\cite{tarasovk3pi}, the structure of the cusp
in the presence of re--summed Coulomb ladders is investigated as well. In case  that
 the same approximation is made, the 
so induced modifications of the decay distribution   agree in the two frameworks.
 Furthermore, pionium production was calculated in
\cite{silagadze}, and compared with data by the NA48/2 collaboration \cite{dilella}.]
\item
Presently, we investigate the emission of real photons \cite{cusprad}. We evaluate the 
infrared divergent contributions in $K\to 3\pi(\gamma)$ in dimensional regularization,
and  organize
 the remaining integrations over phase space in such a manner that the 
 calculations  can be performed numerically 
reasonably fast,
  so that a corresponding FORTRAN code  can be used 
 by the NA48/2 collaboration for performing fits.
\end{itemize}

\noindent

\section{$K_{e4}$ decays}
\subsection{General}
 In the NA48/2 experiment, the general purpose Monte Carlo program
 PHOTOS \cite{photos} is used to calculate electromagnetic corrections.
 In addition, the Sommerfeld factor is applied, to account for the 
Coulomb interaction between charged particles \cite{blochprivate}.

In my talk, I pointed out that in these prescriptions to perform 
radiative corrections, one specific mechanism is not included. Namely, 
the kaon may decay first into a neutral pion pair, that then annihilates 
into two charged pions, or first decay into a charged pion pair, 
that then re--scatters.
In the real world, the neutral pion mass
is smaller than the charged one by about 4.6 MeV\footnote{Throughout the text, 
I use the symbols $M_\pi$ ($M_{\pi^0}$) for the charged
  (neutral) pion mass.}.
  As a result of this, the two contributions to the decay matrix element
 have a different holomorphic structure: the 
neutral (charged) pion loop generates a branch point 
at $s_\pi=4M_{\pi^0}^2$ (at $s=4M_\pi^2$), and  the {\it phase}
of the relevant form factor is affected with a cusp, and does not vanish at
the threshold $s=4M_\pi^2$.
In my talk, I discussed the case of  the scalar form factor $\langle \pi^+\pi^-|\bar u 
u+\bar dd|0\rangle$, where the same effect occurs. Assuming universality, 
I applied the result to $K_{e4}$ decays \cite{internalGR}.
Meanwhile, we have   worked out \cite{CGR} the effect for $K_{e4}$ 
decays as well. I  describe the outcome of the calculation shortly here, and  refer the
 interested reader to our forthcoming publication for details \cite{CGR}.

\subsection{Partial wave expansion: isospin symmetry limit}

The matrix element for $K^+(p)\to \pi^+(p_1)\pi^-(p_2)e^+(p_e)\nu_e(p_\nu)$ is
       \bea
      T = \frac{G_F}{\sqrt{2}} V^\star_{us} \bar{u} (p_\nu) \gamma^\mu
      (1-\gamma_5)  v (p_e) (V_\mu - A_\mu),
       \eea
       where the last factor denotes hadronic matrix elements of the strangeness 
changing  (vector and axial vector) 
      currents,
       \bea
      V_\mu-A_\mu & = & \langle \pi^+ (p_1) \pi^- (p_2) \,\mbox{out}\mid
      (\bar s\gamma_\mu u-\bar s\gamma_\mu\gamma_5 u) \mid K^+ (p)  \rangle\fs
\eea
In the following, I concentrate on the matrix element of the axial vector current, because 
it carries information on the $\pi\pi$ final state interactions and, in particular, on 
the $\pi\pi$ phases. One decomposes $A_\mu$ into Lorentz scalars,
\bea 
     A_\mu  =  -i\frac{1}{M_K} \left [ (p_1+p_2)_\mu F +
      (p_1-p_2)_\mu G + (p_e+p_\nu)_\mu K \right ]\fs
       \eea
      The form factors $F,G, K$  are holomorphic functions of the three
      variables
\bea
s_\pi=(p_1+p_2)^2\scs t=(p_1-p)^2\scs u=(p_2-p)^2\, \fs
\eea
Sometimes, it is useful to use instead
\bea
s_\pi=(p_1+p_2)^2\scs s_\ell=(p_e+p_\nu)^2\scs \cos{\theta_\pi}\scs
\eea
where $\theta_\pi$ is the angle of the $\pi^+$ in the CM system of the two charged pions, 
with respect to the dipion line of flight in the rest system of the kaon 
\cite{cabibbomaksymovicz, blochkaon07}.
In the isospin symmetry limit, one identifies the $\pi\pi$ phases in the matrix element in a
 standard manner, by performing a partial wave expansion, and using unitarity and analyticity, although, in the present case, this is a slightly intricate endeavor \cite{partialwaveexpansion}. 
It is useful to introduce a particular combination of form factors,
\bea
F_1=F+\frac{(M_K^2-s_\pi-s_\ell)\sigma}{\lambda(M_K^2,s_\pi,s_\ell)^{1/2}} \cos{\theta_\pi}G\fs
\eea
Here, $\sigma=\sqrt{1-4M_\pi^2/s_\pi}$, and $\lambda(x,y,z)$ is the triangle
function. The form factor $F_1$ has a simple partial wave expansion,
\bea
F_1=f(s_\pi,s_\ell)+\sum_{k\geq 1}  P_k(\cos{\theta_\pi}) f_k(s_\pi,s_\ell)\fs
\eea
In the low--energy region
$s_\pi\leq 16 M_\pi^2$,  $f_k$ carry the $\pi\pi$ phases \cite{partialwaveexpansion} 
in the pertinent isospin channel. In the following, I consider the lowest partial wave
 $f(s_\pi,s_\ell)$.
In the  interval $4M_\pi^2\leq s_\pi\leq 16 M_\pi^2$, its phase coincides with the isospin 
zero $S$-wave phase $\delta_0^0$ in elastic $\pi\pi$ scattering,
\bean\label{eq:fphase}
f_+=e^{2i\delta_0^0}f_-,\,f_\pm=f(s_\pi\pm i\epsilon,s_\ell)\fs
\eean
\begin{figure}[t]
\begin{center}
\includegraphics[width=13cm]{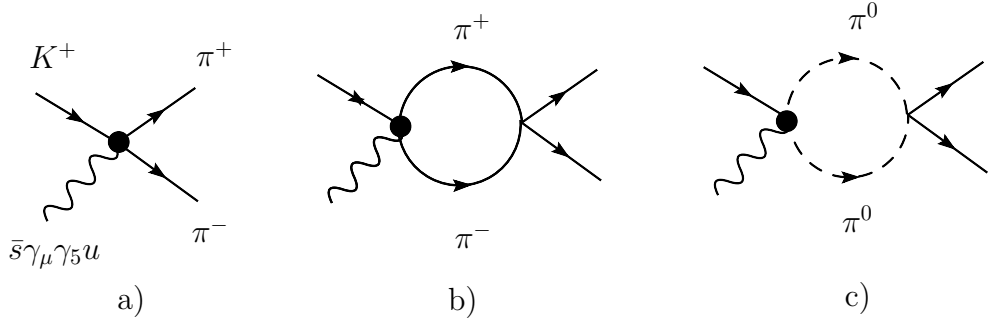}
\end{center}
\caption{Some of the graphs that contribute to the matrix element of the axial current 
at tree and one--loop order.
The filled vertex indicates that the axial current also couples to a single
kaon line. That graph contributes to the form factor $K$. There are many 
additional graphs at one--loop order, not displayed in the figure.}
\label{fig:treeoneloop}
\end{figure}
It is instructive to calculate the form factors in chiral perturbation theory and to verify that 
$F_1$ indeed has the behaviour just discussed. For this, it is sufficient to consider the 
effective Lagrangian
\bea
{\cal L}_2=\frac{F_0^2}{4}\langle D_\mu U D^\mu U^\dagger +2B_0{\cal M}(U+U^\dagger)\rangle\,,
\eea
where the covariant derivative $D_\mu U$ contains the external vector and axial vector 
currents, and ${\cal M}=\mathrm{diag}(\hat m,\hat m,m_s)$. 
Some of the graphs that contribute at tree level and at one loop are displayed in 
Figure \ref{fig:treeoneloop}. The result is \cite{kl4oneloop}
\bea
f(s_\pi,s_\ell)=\frac{M_K}{\sqrt{2}F_0}\left\{1+\Delta(s_\pi)+H(s_\pi,s_\ell)+O(p^4)\right\}\,,
\eea
with
\bea
\Delta(s_\pi)&=&\frac{1}{2F_0^2}(2s_\pi-M_\pi^2){\bar J} (s_\pi)\,,\nnnl
16\pi^2\bar J(s_\pi)&=&\sigma\left(\ln{\frac{1-\sigma}{1+\sigma}}+i\pi\right) +2\,, \quad s_\pi \geq 4M_\pi^2\fs
\eea
Here, $M_\pi$ ($F_0$) denotes the pion mass (pion decay constant), at leading order in the chiral expansion.
The quantity $H(s_\pi,s_\ell)$ is real in the interval of elastic $\pi\pi$ scattering.
It is now seen that  $f$ indeed has the property (\ref{eq:fphase}) at this 
order in the low--energy expansion,
 with
\bean\label{eq:phasesymm}
\delta_0^0=\frac{(2s_\pi-M_\pi^2)}{32\pi F_0^2}\sigma\,.
\eean
This is the phase of the  isospin zero $S$-wave, in tree approximation.
\subsection{Partial wave expansion: the real world}
In reality, experiments are not carried out in the paradise world just discussed: we have not 
included so far photons, nor did we consider isospin breaking effects generated by  different up and 
down quark masses. Here, I investigate these effects in several steps:
\begin{enumerate}
\item[i)]
I assume that the manner in which real and virtual photons
are treated in the analysis of the NA48/2 experiment (PHOTOS + Sommerfeld factor) is a decent approximation
to the effects generated by soft photons.
\item[ii)]
This procedure  misses the  effects generated by the pion and kaon
mass differences, and by the  quark mass difference $m_d-m_u$. 
These must therefore be taken into account separately.
\item[iii)]
ChPT is the appropriate tool to evaluate these contributions.
\item[iv)]
I assume that PHOTOS+Sommerfeld factor + mass effects provide a good approximation to 
the full isospin breaking contributions.
\end{enumerate}
\noindent
\underline{Remark:} One may envisage a more ambitious procedure \cite{knecht}, by working 
out the relevant matrix elements in the framework of ChPT including photons and leptons \cite{chptlept}, 
and then constructing  a new event generator, to be used in the analysis of $K_{e4}$ decays.
 [A one--loop calculation  was already performed in \cite{cuplov}. It needs to be
 checked, and brought into a form which is suitable for the present purpose.] Eventually, such an analysis  
might lead to an improved  algorithm,  but I consider it  a long term project.
\underline{End of remark.}

According to iii), we simply need to perform a ChPT calculation of the effects 
generated by the mass differences. This is rather easy to achieve at one--loop order: one adapts 
the quark mass matrix, ${\cal M}\to \mathrm{diag}(m_u,m_d,m_s)$, and enlarges
 the 
Lagrangian ${\cal L}_2$ \cite{urech},
\bean\label{eq:Lbreaking}
{\cal L}_2\to {\cal L}_2+C\langle QUQU^\dagger\rangle\,,\,Q=\frac{e}{3}\mathrm{diag}(2,-1,-1)\,,
\eean
where $C$ is a new low--energy constant, that breaks the isospin symmetry of 
the meson masses: $M_\pi\neq M_{\pi^0},M_K\neq M_{K^0}$.

\begin{figure}[t]
\begin{center}
\includegraphics[width=4cm]{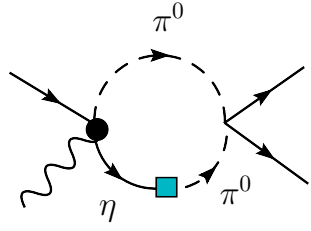}
\end{center}
\caption{The contribution from $\pi^0-\eta$ mixing, at leading order in 
$m_d-m_u$.The filled square denotes the vertex 
from $\pi^0-\eta$ mixing. Otherwise,
the notation is the same as in Figure 1.}
\label{fig:mixing}
\end{figure}

The effect of the replacement (\ref{eq:Lbreaking}) is twofold: first, as just mentioned, 
the meson masses split. As a result of this, the loop contributions
in Fig.~\ref{fig:treeoneloop}b),c) have a different threshold, and the phase of the form factor $f$ 
generates a cusp. 
 Second, in addition to the  graphs displayed in 
Figure \ref{fig:treeoneloop}, there is a new contribution shown in Figure \ref{fig:mixing}: the kaon interacts with the axial current to generate a $\pi^0\eta$ intermediate state. Because $m_u\neq m_d$, the $\eta$ can transform back into a neutral pion, that then re--scatters with the second neutral pion into a charged pion pair.

Working out the relevant diagrams, one finds that the phase (\ref{eq:phasesymm}) becomes
 in the elastic region
\bean\label{eq:phase}
\delta_0^0\to \delta=\frac{1}{32\pi F_0^2}\left\{(4\Delta_\pi +s_\pi)\sigma 
+(s_\pi-M_{\pi^0}^2)\left(1+\frac{3}{2R}\right)\sigma_0\right\}\, ,
\eean
with
\bea
\Delta_\pi=M_\pi^2-M_{\pi^0}^2\,,\quad \sigma_0=\sqrt{1-4M_{\pi^0}^2/s_\pi}\,,
\quad R=\frac{m_s-\hat m}{m_d-m_u}\fs
\eea
The one--loop expressions for the form factors $F,G$
given in Refs.~\cite{cuplov}  contain the effects considered here,
up to terms of order $\alpha_{QED}(m_d-m_u)$. 
 
\begin{figure}[h]
\begin{center}
\epsfig{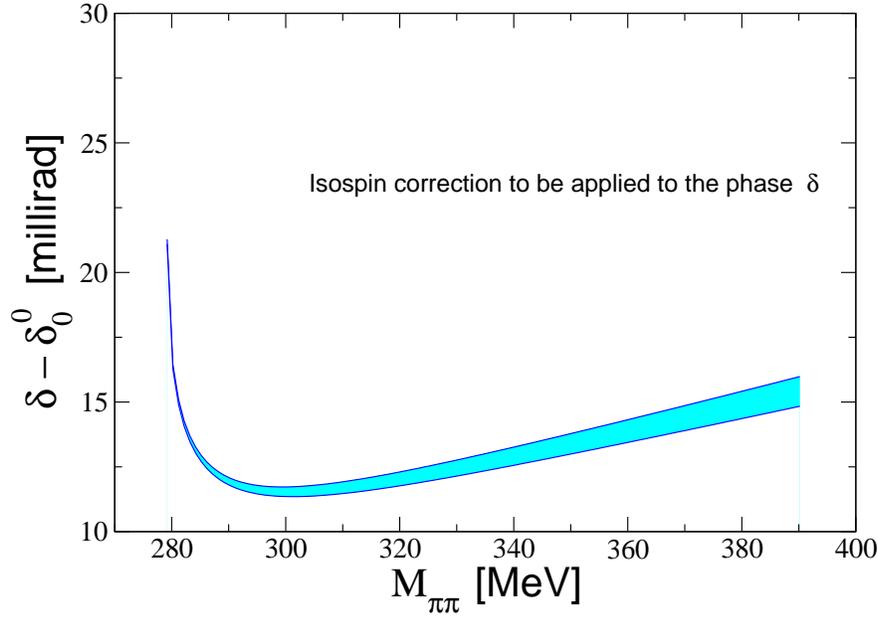}
\end{center}
\caption{The isospin breaking correction that must be subtracted from the phase $\delta$ 
measured in $K_{e4}$ decays. The width of the band reflects the uncertainty in the ratio $R$.
 \label{fig:correction}}
\end{figure}

I consider the result (\ref{eq:phase}) to be very interesting, for the following reasons. 
First, due to the presence of the phase space factor $\sigma_0$, the phase $\delta$ does not vanish at the threshold $s_\pi=4M_\pi^2$. 
 This unexpected behaviour of the phase is the cusp effect already 
experienced in $K\to 3\pi$ decays, 
 with the role of neutral and charged pions interchanged. 
 Second, the difference $\delta-\delta_0^0$ is positive for $s_\pi$ above the threshold,
and even {\it increases} at large $s_\pi$,
\bea
\delta-\delta_0^0&=&\frac{3s_\pi}{64\pi F_0^2}\frac{1}{R}+O(1)\,,\quad s_\pi/M_\pi^2 \gg 1\fs
\eea

We now come to the main point. According to point iv) above, it is the phase $\delta$ that is measured 
in $K_{e4}$ decays (up to contributions from higher orders in the chiral expansion). Therefore, before
comparing the phase so determined with ChPT predictions, one has to subtract from the measured phase 
the (positive) difference $\delta -\delta_0^0$, because $\delta_0^0=\delta - (\delta-\delta_0^0)$.
 In Figure \ref{fig:correction} we display this difference  
 in the relevant decay region, for $R=37\pm4$\footnote{This value for $R$ should be considered as 
preliminary -- it was used in my talk for illustrative purposes. 
 A more refined estimate will be provided in \cite{CGR}. Of course, the conclusions to be drawn from 
the isospin breaking effects considered here will not change.}. The width of the band reflects 
the uncertainty in $R$. 
 [As I pointed out in my talk, two--loop contributions are modest in the 
analogous case of the scalar form factor of the pion \cite{CGR}.]
It is seen that the isospin 
correction  is quite substantial -- 
well above the  uncertainties quoted for the measured
 phase \cite{blochkaon07}. 
  [In Ref.~\cite{tarasovke4}, the cusp in $K_{e4}$
 decays was investigated as well. The expressions presented there do not agree with
 (\ref{eq:phase}), because these authors do not take into account derivative couplings of
 the $\pi\pi$ amplitude, as is dictated by chiral symmetry.] 

Colangelo has performed fits to $K_{e4}$ data, with and without  isospin breaking corrections
applied. It turns out that  the former discrepancy \cite{na48old} 
of the NA48/2 data  with the prediction \cite{scattnpb} disappears, once isospin breaking effects are taken 
into account in the manner just described, see \cite{colangelokaon07}. 
 Colangelo also  shows that the former 
agreement between the chiral prediction and the E865 data \cite{ke4}  becomes marginal. 
This is  an issue that should be understood, because it is independent of the
special effects considered here.
 On the other hand, since the NA48/2 data are so precise, they will dominate the
 E865 result, in any case.

\hspace{5cm}{.}

{{{\bf Note added:} Isidori has meanwhile 
worked out soft photon corrections for $K^+\to\pi^+\pi^+\pi^-$ \cite{isidorirad}.}}

\vskip5mm

{\bf Acknowledgments}

It is a pleasure to thank the organizers for the invitation to give this talk, and for the excellent 
organization of this exciting conference at that beautiful place. 
I thank M.~Bissegger, G.~Colangelo, A.~Fuhrer, B.~Kubis, 
H.~Leutwyler and A.~Rusetsky for a most enjoyable collaboration,   B.~Bloch--Devaux,
L.~DiLella and I.~Mannelli for  explaining to me with patience 
details of the  data analysis. I am grateful to   M.~Knecht for discussions, 
 and A.V.~Tarasov for  
 communications concerning topics considered here.
Finally, I thank G.~Colangelo, H.~Leutwyler, J.~R.~Pel\'aez, A.~Rusetsky and F.~J.~Yndur\'ain 
for  useful comments on the manuscript.

\end{document}